\newcommand{\diracslash}[1]{#1\llap{/\kern2pt}}

\newcommand{\be}{\begin{equation}}
\newcommand{\ee}{\end{equation}}
\newcommand{\bea}{\begin{eqnarray}}
\newcommand{\eea}{\end{eqnarray}}
\newcommand{\ba}[1]{\begin{array}{#1}}
\newcommand{\ea}{\end{array}}

\documentclass[amssymb,nobibnotes,nofootinbib,aps,pra,showpacs]{revtex4}
\usepackage{graphicx}
\begin{document}
\setlength{\topmargin}{-0.05in}

\title{Probing pairing gap in Fermi  atoms by light scattering}

\author{Bimalendu Deb}
 \affiliation{ Physical
Research Laboratory, Navrangpura, Ahmedabad 380 009, India}

\date{\today}

\begin{abstract}
We study stimulated scattering of polarized light in a
two-component Fermi gas of atoms  at zero temperature. Within the
framework of Nambu-Gorkov formalism, we calculate the response
function of superfluid gas taking into account the final state
interactions. The dynamic structure factor deduced from the
response function provides information about  the pairing gap and
the momentum distributions of  atoms. Model calculations using
local density approximation indicates that the pairing gap of
trapped Fermi gas may be detectable by Bragg spectroscopy due to
stimulated scattering.
\end{abstract}

\def\be{\begin{equation}}
\def\ee{\end{equation}}
\def\bearr{\begin{eqnarray}}
\def\eearr{\end{eqnarray}}
\def\zbf#1{{\bf {#1}}}
\def\bfm#1{\mbox{\boldmath $#1$}}
\def\hf{\frac{1}{2}}

\pacs{03.75.Fi,74.20.-z,32.80.Lg}

\maketitle

\section{Introduction}

Since the first realization of quantum degeneracy in an atomic
Fermi gas by Jin's group \cite{jin} in 1999, cold Fermi atoms have
been in focus of research interest in current physics. In a series
of remarkable experiments, several groups
\cite{hulet,solomon,thomas,mit,italy,grimm} have demonstrated many
new aspects of cold degenerate atomic Fermi gases.  The ability to
change interatomic  interaction ranging from strong attraction to
strong repulsion makes Fermi atoms the most favorable laboratory
system for testing theoretical models in diverse fields. In
particular, research with Fermi atoms has relevance in the field
of superconductivity \cite{pred1,pred2}. The basic mechanism
behind superconductivity and Fermi superfluidity is
particle-particle pairing with an energy gap. Recently, two
groups-Innsbruck \cite{gap1} and JILA \cite{gap2} have
independently reported the measurement of pairing gap in Fermi
atoms.  Furthermore, two groups-Duke and Innsbruck
\cite{duke,innsbruck} have measured collective oscillations
 which indicate the
occurrence of fermionic superfluidity \cite{stringari} in atomic
Fermi gas. The superfluid pairing is believed to occur near the
crossover \cite{nozrink,randeria,crossover} between the predicted
BCS state of  atoms and the Bose-Einstein condensation of
molecules formed from Fermi atoms. Several groups have produced
Bose-Einstein condensates (BEC) \cite{molecules} of molecules
formed from degenerate Fermi atoms. Several other recent
experimental \cite{expt} and theoretical investigations
\cite{theory} have revealed many intriguing aspects of cold Fermi
gases.

Several theoretical proposals \cite{zoller,huletp} have been made
for probing pairing gap. A method has been suggested to use
resonant light \cite{zoller} to make an interface between normal
and superfluid atoms. This has been recently implemented
\cite{gap1,theogap1}, albeit in the radio frequency domain. A
number of authors \cite{mottelson,griffin,minguzzi} have
theoretically investigated Bogoliubov-Anderson (BA)  mode
\cite{bamode,anderson,martin} in fermionic atoms as a signature of
superfluidity.  BA mode constitutes a distinctive feature of
superfluidity in neutral Fermi systems since it is associated with
long wave  Cooper-pair density fluctuations.

Our purpose here is to study Bragg spectroscopy with off-resonant
polarized lasers as a method for detecting the paring gap. Bragg
spectroscopy has been used by Ketterle's group for measuring
structure factor of an atomic BEC \cite{bragg}. Bragg scattering
in superfluid  Fermi atoms has an analogy with Raman scattering in
electronic superconductors \cite{abrikosov}.  In the next section
we describe polarization-selective light scattering. We then
discuss briefly the method of calculation of response function of
superfluid Fermi gas using Green function techniques
\cite{gorkov,nambu}.  We present the main results which suggest
the possibility of detecting pairing gap by scattering of
circularly polarized light.

\section{polarization-selective light scattering}

When two off-resonant laser beams with a small frequency
difference are impinged on atoms, the scattering of one laser
photon is stimulated by the other photon. In this process, one
laser photon  is annihilated and reappeared as a scattered photon
propagating along  the other laser beam. The magnitude of momentum
transfer is $q \simeq 2 k_L \sin(\theta/2) $, where $\theta$ is
the angle between the two beams and $k_L$ is the momentum of a
laser photon. To illustrate the main idea, we specifically
consider trapped $^6$Li Fermi atoms in their two lowest hyperfine
spin states $\mid g_1 \rangle = \mid 2{\rm S}_{1/2}, F = 1/2,
m_{F} = 1/2 \rangle$ and $\mid g_2 \rangle = \mid 2{\rm S}_{1/2},
F=1/2, m_F = -1/2 \rangle $.  For simplicity, we consider that the
number of atoms in each spin component is the same.  An applied
magnetic field tuned near the Feshbach resonance ($\sim 850$
Gauss) results in splitting between the two spin states by  $\sim
75$ MHz \cite{zwierlein}, while the corresponding splitting
between the excited states $ \mid e_1 \rangle = \mid 2{\rm
P}_{3/2}, F=3/2, m_{F} = -1/2 \rangle $ and $\mid e_2\rangle =
\mid 2{\rm P}_{3/2}, F = 3/2, m_{F} = -3/2 \rangle $ is $\sim 994$
MHz \cite{thomas}. Let both the laser beams be $\sigma_{-}$
polarized and tuned near the transition $\mid g_2\rangle
\rightarrow \mid e_2\rangle$. Then the transition between the
states $\mid g_1 \rangle $ and $\mid e_2 \rangle $ would be
forbidden while the transition $\mid g_1 \rangle \rightarrow \mid
e_1\rangle $ will be suppressed due to the large detuning $\sim
900$ MHz. This leads to a situation where the Bragg-scattered
atoms remain in the same initial internal state $\mid e_2\rangle$.
Similarly, atoms in state $\mid g_1 \rangle$ only would undergo
Bragg scattering when two $ \sigma_{+}$ polarized lasers are tuned
near the transition $\mid g_1 \rangle \rightarrow \mid 2{\rm
P}_{3/2}, F = 3/2, m_F = 3/2\rangle$. Thus, we infer that in the
presence of a high magnetic field,  it is possible to scatter
atoms selectively of either spin components only by using
circularly polarized Bragg lasers. We assume that both the laser
beams are $\sigma_-$ polarized and tuned  near the transition
$\mid g_2\rangle \rightarrow \mid e_2 \rangle $. Under such
conditions, considering a uniform gas of atoms, the effective
laser-atom interaction Hamiltonian in electric-dipole
approximation can be written as
 \bearr
 H_I = \hbar \Omega \sum_{{\mathbf k},\sigma=1,2}\gamma_{\sigma}
  \hat{c}_{\sigma}^{\dagger}({\mathbf k }+ {\mathbf q})
  \hat{c}_{\sigma}({\mathbf
  k})+
 {\mathrm H.c.} \label{eq2}
  \eearr
where $\hat{c}_{\sigma}(k)$ represents annihilation operator of an
atom with momentum $\mathbf{k}$ in the internal state $\sigma$.
The subscript $1(2)$ refer to the state $\mid g_1\rangle$ ($\mid
g_2\rangle$), $\Omega = (\Omega_1+\Omega_2)/2$,  and $\gamma_i =
\Omega_i/\Omega$. Here $\Omega_i$ denotes the two-photon Rabi
frequency for the transitions $\mid g_i\rangle \rightarrow \mid
e_i\rangle \rightarrow \mid g_i\rangle$. For both the laser beams
having $\sigma_{-}$ polarization tuned near $\mid g_2\rangle
\rightarrow \mid e_2 \rangle $, we have $\Omega_2 >\!> \Omega_1$.
One can identify the operator $\hat{\rho}_{\sigma}^{(0)}({\mathbf
q}) = \sum_{{\mathbf k}} \hat{c}_{\sigma}^{\dagger}({\mathbf{k +
q}}) \hat{c}_{\sigma}({\mathbf k})$ as the Fourier transform of
the density operator.

\section{the formalism}
The scattering probability is given by the susceptibility \bearr
\chi(\mathbf{q},\tau-\tau') = -\langle
T_{\tau}[\rho_q^{(\gamma)}(\tau)\rho_{-q}^{(\gamma)}(\tau')]\rangle.
\eearr where $
 \rho_q^{(\gamma)} = \sum_{k,\sigma} \gamma_{\sigma}
 a_{k+q,\sigma}^{\dagger}a_{k,\sigma}
 $,
$T_{\tau}$ is the complex time $\tau$ ordering operator and
$\langle \cdots \rangle$ means thermal averaging. The dynamic
structure factor is related to $\chi$ by
$\chi(\mathbf{q},\omega_n)$ as \bearr S(\mathbf{q},\omega) = -
\frac{1}{\pi}[1+n_B(\omega)]\rm{Im}[ \chi(\mathbf{q},z=\omega +
i0^{+})]. \label{flucdiss}\eearr This follows from generalized
fluctuation-dissipation theorem. In order to treat collective
excitations, it is essential to go beyond Hartree approximation
and apply either a kinetic equation or a time-dependent
Hartree-Fock equation  or a random phase approximation
\cite{martin}. The essential idea is to take into account the
residual terms which are neglected in the BCS approximation and
thereby treat the off-diagonal matrix elements (vertex functions)
of single-particle operators in a more accurate way
\cite{schrieffer,martin}.

To study light scattering in Cooper-paired fermionic atoms, we
apply Nambu-Gor'kov formalism \cite{nambu,gorkov} of
superconductivity \cite{schrieffer}. Using the familiar Pauli
matrices, the susceptibility can be expressed as \bearr
\chi(\mathbf{q},\omega) = \int
\frac{d^4k}{(2\pi)^4i}\rm{Tr}[\tilde{\gamma}_k\mathbf{G}(k_{+})\Gamma(k_+,k_-)\mathbf{G}(k_-)]
\eearr where the Green function has a  matrix form as \bearr
 G(k) = \frac{k_0 \tau_{0} + \xi_k\tau_3 +
 \Delta_k\tau_1}{k_0^2 - E_k^2 + i\delta}, \label{green}\eearr where $E_k =
 \sqrt{\xi_k^2 + \Delta_k^2}$ and $\xi_k = \epsilon_k-\mu$ with $\epsilon_k = \hbar^2
 k^2/(2m)$. Here $\tau_0$ is a $2\times 2$ unit matrix.
The vertex equation is \bearr \Gamma(k_+,k_-) = \tilde{\gamma} + i
\int \frac{d^4k'}{(2\pi)^4} \tau_3 \mathbf{G}(k_+')
\Gamma(k_+',k_-')\mathbf{G}(k_-')\tau_3V(\mathbf{k},\mathbf{k}'),
\label{vertex} \eearr where $k_{\pm} = k \pm q/2$ and $k =
(\mathbf{k},k_0)$ is the
 energy-momentum  4-vector whose components are $k_3 = \xi_k$
 and $k_4=ik_0$.  The bare vertex is a diagonal matrix:
 $\tilde{\gamma} = \rm{Diag}.[
 \gamma_{1},  -\gamma_{2}] $.  Using Pauli matrices $\tau_0$ and
$\tau_3$, this can be rewritten as $\tilde{\gamma} = \gamma_0(k)
\tau_0 + \gamma_3(k) \tau_3$, where $\gamma_0 = [\gamma_{1} -
\gamma_{2}]/2$ and $\gamma_3 = [\gamma_{1} + \gamma_{2}]/2$. For
unpolarized light in the absence of magnetic field, $\gamma_{1} =
\gamma_{2}$.  However, if the incident light is polarized,
$\gamma_{1} \ne \gamma_{2}$. In the specific case of $\sigma_{-}$
polarization as discussed above, we have $\gamma_3 \simeq
-\gamma_0 \simeq \gamma_{2}/2$. If the  potential
$V(\mathbf{k},\mathbf{k}')$ is separable in $\mathbf{k}$ and
$\mathbf{k}'$, then Eq. (\ref{vertex}) is analytically solvable.
We replace $V(\mathbf{k},\mathbf{k}')$ by the  potential
$V=4\pi\hbar^2a_s/(2\tilde{m})$, ($\tilde{m}=m/2$ being the
reduced mass) expressed in terms of s-wave scattering length
$a_s$. The strong-coupling limit may be accessed by first
renormalizing the BCS mean-filed interaction and then taking the
limit $a_s \rightarrow \pm \infty$ as we will discuss later. The
four-dimensional integrals of Eq. (\ref{vertex}) can be performed
following the established method of relativistic quantum
electrodynamics  as applied for studying collective excitations in
a superconductor \cite{vaks}. The detailed method of solution  is
discussed  elsewhere \cite{deb2}. We here present the final result
 \bearr \chi(\mathbf{q},\omega) =
2N(0)\gamma_0^2\langle B \rangle + 2 N(0)   \left [\langle A
\rangle + \frac{\omega^2 \langle f \rangle ^2 }{4\Delta^2\langle
\beta^2f\rangle}\right ] \gamma_3^2 \label{chinew}\eearr where
$N(0)$ is the density of states at the Fermi surface and
 \bearr A =
\frac{(\mathbf{v}_k.\mathbf{p}_q)^2-\omega^2
f}{\omega^2-(\mathbf{v}_k.\mathbf{p}_q)^2}, \hspace{0.5cm}
 B =
\frac{(\mathbf{v}_k.\mathbf{p}_q)^2(1-f)}{\omega^2-(\mathbf{v}_k.\mathbf{p}_q)^2}.
\eearr Here $\mathbf{p}_q = \hbar \mathbf{q}$ and $\mathbf{v}_k$
is the velocity of the atoms with momentum $\mathbf{k}$ and \bearr
f(q) = \frac{\arcsin \beta}{\beta\sqrt{1-\beta^2}}, \hspace{0.5cm}
 \beta^2 = \frac{\omega^2 -
(\mathbf{v}_k.\mathbf{p}_q)^2}{4\Delta^2}. \eearr The symbol
$\langle X \rangle$ implies averaging  of a function $X$ over the
chemical potential surface:  $ \langle X \rangle = [N(0)]^{-1}\int
d^3\mathbf{k}\delta(\epsilon_k)  X$. As $\omega \rightarrow \omega
+ i 0^+$, $\beta \rightarrow \beta + i 0^+$,  we have the
following analytic property of $f(\beta)$:
 \bearr f(\beta) &=& -\frac{{\rm arcsinh} \sqrt{\beta^2-1}}{\beta
\sqrt{\beta^2-1}} + \frac{i\pi/2}{\beta\sqrt{\beta^2-1}},
 \hspace{0.2cm} \beta > 1 \label{anal} \eearr

\section{Strong-coupling limit}

To access the strong-coupling limit, the chemical potential $\mu$
and the gap $\Delta_k$ should  be obtained by solving the gap
equation \bearr \frac{m}{4\pi\hbar^2a_s} = \frac{1}{V}
\sum_{{\mathbf k}} \left (\frac{1}{2\epsilon_k} -
\frac{1}{2E_k}\right ) \label{gap} \eearr along with the equation
\bearr n= \frac{1}{6\pi^2} k_F^3 = \frac{1}{V} \sum_{{\mathbf k}}
\left (1- \frac{\xi_k}{E_k}\right ). \label{num} \eearr of the
density of single component.  Note that the eq. (\ref{gap}) is
obtained by regularizing the zero-temperature BCS gap equation
with a mean-field parameterized by two-body scattering length  as
done in Ref. \cite{randeria}. This approach fails to account for
pairing fluctuation effects which are particularly significant
near $T_c$ in strong-coupling regime. However, far below $T_c$,
the correction due to the pairing fluctuation is very small
\cite{randeria}.  Based on this regularized mean-field approach
and local density approximation (LDA), the zero-temperature
density profiles \cite{strinati}, momentum distribution
\cite{pitaevskii} and  the finite temperature effects
\cite{perali} of superfluid trapped Fermi atoms have  been
recently studied . The two coupled eqs. (\ref{gap}) and
(\ref{num}) admit analytical solutions \cite{analyt}. In the
unitarity limit the solutions yield $\Delta \simeq 1.16\mu$, $\mu
= (1+\beta)\epsilon_F$, where $\beta = -0.41$
\cite{analyt,pitaevskii} is a constant. Many other recent
theoretical \cite{phand,heiselberg} and experimental
\cite{solomon,barten} studies have established the universality of
Fermi gas in the unitarity limit.

\begin{figure}
\includegraphics[width=3.25in]{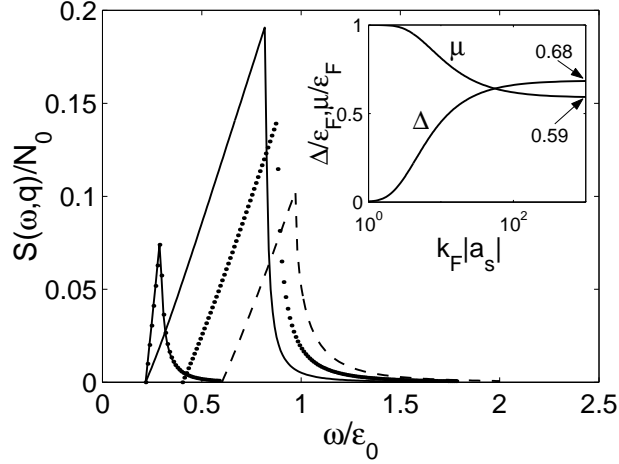}
 \caption{Dimensionless dynamic structure factor $S(\delta,{\mathbf
 q})/N(0)$ of a uniform superfluid Fermi gas
 is plotted as a function of dimensionless energy
 transfer $\omega/\epsilon_0$ ($\epsilon_0$ is the Fermi energy)
 for different values of the scattering length
 $a_s= 2.76 k_F^{-1}$ (solid),  $a_s= 3.89 k_F^{-1}$ (dotted), $a_s= 5.47 k_F^{-1}$ (dashed)
 for a fixed momentum transfer $q = 0.8 k_F$. The dash-dotted
 curve is plotted for $a_s= 2.76k_F^{-1}$ and $q=0.4 k_F$. The inset shows the
 variation of the gap $\Delta$ and the chemical potential $\mu$ as a function
 of $a_s$.}
 \label{figdel}
 \end{figure}

\begin{figure}
 \includegraphics[width=3.25in]{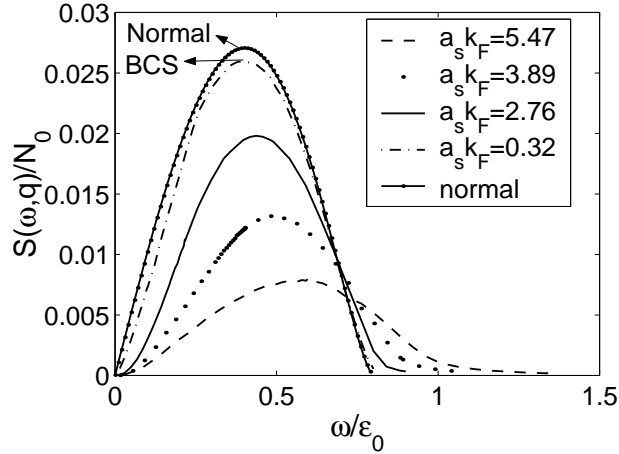}
 \caption{Same as in Fig. (\ref{figdel}) but for a trapped  superfluid Fermi gas
 for a fixed momentum transfer $q = 0.8 k_F$. Here $\epsilon_0$ is the Fermi energy at the trap center}
 \label{figdel}
 \end{figure}

\section{Leading approximations}
  With the use of Eqs. (\ref{chinew},\ref{anal}) in Eq. (\ref{flucdiss}), the dynamic
structure factor in the leading approximation in terms of
$\beta^{-1}$ can be written as \bearr S(\mathbf{q},\omega) =N(0)
 \frac{1}{4\Delta^2}\left [\gamma_3^2\left \langle
\frac{\omega^2}{\beta^3\sqrt{\beta^2-1}} \right \rangle +
\gamma_0^2 \left \langle
\frac{(\mathbf{p}_q.\mathbf{v}_k)^2}{\beta^3\sqrt{\beta^2-1}}
\right \rangle \right ], \hspace{0.2cm} \beta >1
\label{leadapp}\eearr This is also obtainable from  the BCS-
Bogoliubov mean-filed treatment as shown in \cite{deb2}. Although
this leading approximation  takes into account excitations in the
particle-hole continuum, it fails to account for the in-gap
collective modes. It can be verified \cite{deb2} that in the
regime of large momentum and energy transfer ($\beta
>\!>1$), the dynamic structure factor of Eq. (\ref{leadapp})
approximately satisfies the f-sum rule \bearr \int \omega
S(\mathbf{q},\omega)d\omega \simeq \frac{Nq^2}{2m}, \hspace{0.4cm}
\xi q >\!>1 \eearr where $N$ is the total number of particles.

Now let us consider the case $0 \le \beta <\!< 1$, that is
$\mathbf{v}_k.\mathbf{p}_q \le \omega <\!< 2\Delta$. In this case,
the second term in the coefficient of $\gamma_3^2$ in Eq.
(\ref{chinew}) dominates over all other terms. This term leads to
BA  mode appearing as a pole of $\chi$. BA mode restores the
continuous symmetry  that is broken by BCS ground state. In the
limit $q \rightarrow 0$ and $\omega \rightarrow 0$, $f \simeq 1$
and
 hence the pole is \bearr \omega_{\rm{BA}} =
\frac{1}{\sqrt{3}}v_Fp_q .\eearr  In the low momentum and low
energy limit ($0 \le \beta <\!< 1$)
 the dynamic structure factor can be obtained by
linearizing  the denominator of the second term in Eq.
(\ref{chinew}) around the BA mode. By approximating $f \simeq 1$,
we then obtain \bearr S(\mathbf{q},\omega) = N(0)\gamma_3^2
\frac{\omega^2}{2\omega_{BA}}\delta(\omega - \omega_{BA}). \eearr
With $\gamma_3\rightarrow1$,  this satisfies the $f-$sum rule.
 BA mode is well defined in the low momentum
regime, i.e., for $\xi q=v_Fp_q/(2\Delta) <\!<1$. For large
momentum, it becomes ill defined due to Landau damping. Ohashi and
Griffin \cite{griffin} have provided a detailed theoretical
treatment of this mode in the BCS-BEC crossover in Fermi atoms.
Minguzi {\it et. al.} \cite{minguzzi} have found that this mode
appears as a prominent asymmetric peak in the spectrum of density
fluctuation at a very low momentum and energy.

\section{Results and discussions}
Figure 1 and 2 show $S(\omega,{\mathbf q})$   as a function of
$\omega$ for a uniform and trapped gas, respectively, for
different values of $a_s$. In the case of trapped gas, we use LDA
with local chemical potential $\mu ({\mathbf r})$ determined from
equation of state of interacting Fermi atoms in a harmonic trap.
When $a_s$ is large, the behavior of $S(\delta,{\mathbf q})$ is
quite different from that of normal as well as weak-coupling BCS
superfluid. This can be attributed to the occurrence of large gap
for large $a_s$.  In contrast to the case of a uniform superfluid
\cite{abrikosov}, $S(\delta,{\mathbf q})$ for a superfluid trapped
Fermi gas has a structure below $2\Delta(0)$, where $\Delta(0)$ is
the gap at the trap center. As the energy transfer decreases below
$2\Delta(0)$, the slope of $S(\delta,{\mathbf q})$ gradually
reduces. Particularly distinguishing feature of $S(\delta,{\mathbf
q})$ of a superfluid compared to normal fluid is gradual  shift of
the peak as $a_s$ or $\Delta$ increases. The quasiparticle
excitations occur only when  $2\Delta({\mathbf x}) < \omega$. This
implies that, when $\omega$ is less than $2\Delta(0)$, the atoms
at the central region of the trap can not contribute to
quasiparticle response.

\section{conclusions}

 Order of magnitude analysis of Ref. \cite{deb1}
suggests that, with large momentum transfer, it may be possible to
distinguish the scattered atoms in time of flight images. A
comparison of images with and without Bragg pulses may reveal
information about the momentum and density distribution of the
scattered atoms. Furthermore, the polarization-selective Bragg
spectroscopy  may lead to better precision in time-of-flight
spin-selective measurements \cite{molecules,colorado} since they
will be in the same spin component. It is possible to select
counter propagating  scattered atoms using three or four beam
scattering configurations \cite{deb1}. One can then explore the
possibility of measuring the correlation of two scattered atoms
with opposite momentum by the technique as used in recent studies
\cite{theoshot,jinshot}.

\acknowledgments{The author is thankful to P. K. Panigrahi and H.
Mishra for helpful discussions.}

\end{document}